\documentclass[aps,pra,twocolumn,showpacs,superscriptaddress,english]{revtex4-1}
\usepackage{amssymb}
\usepackage{graphicx}
\usepackage{amsmath}
\usepackage{amsthm}
\usepackage{amsfonts}
\usepackage{txfontsb}
\usepackage{color}
\usepackage{bm}
\usepackage{bbm}
\usepackage{url}
\usepackage[driverfallback=dvipdfm]{hyperref}
\hypersetup{pdfpagemode=FullScreen,colorlinks=true,breaklinks,urlcolor=blue,linkcolor=blue,citecolor=blue}

\makeatletter
\def\iddots{\mathinner{\mkern1mu\raise\p@
    \hbox{.}\mkern2mu\raise4\p@\hbox{.}\mkern2mu
    \raise7\p@\vbox{\kern7\p@\hbox{.}}\mkern1mu}}
\def\adots{\mathinner{\mkern2mu\raise\p@\hbox{.} 
 \mkern2mu\raise4\p@\hbox{.}\mkern1mu
 \raise7\p@\vbox{\kern7\p@\hbox{.}}\mkern1mu}}
\makeatother

\begin{document}

\global\long\def\id{\mathbbm{1}}
\global\long\def\ui{\mathbbm{i}}
\global\long\def\ud{\mathrm{d}}

\title{Signal detection in nearly continuous spectra and symmetry breaking}

\author{Vincent Lahoche} \email{vincent.lahoche@cea.fr}   
\affiliation{Université Paris-Saclay, CEA, List, F-91120, Palaiseau, France}

\author{Dine Ousmane Samary}
\email{dine.ousmanesamary@cipma.uac.bj}
\affiliation{Université Paris-Saclay, CEA, List, F-91120, Palaiseau, France}
\affiliation{International Chair in Mathematical Physics and Applications (ICMPA-UNESCO Chair), University of Abomey-Calavi,
072B.P.50, Cotonou, Republic of Benin}

\author{Mohamed Tamaazousti}\email{mohamed.tamaazousti@cea.fr}\affiliation{Université Paris-Saclay, CEA, List, F-91120, Palaiseau, France}


\begin{abstract}
The large scale behavior of systems having a large number of interacting degrees of freedom is suitably described using renormalization group, from non-Gaussian distributions. Renormalization group techniques used in physics are then expected to be helpful for issues when standard methods in data analysis break down. Signal detection and recognition for covariance matrices having nearly continuous spectra is currently an open issue in data science and machine learning. Using the field theoretical embedding introduced in arXiv:2011.02376 to reproduces experimental correlations, we show in this paper that the presence of a signal may be characterized by a phase transition with $\mathbb{Z}_2$-symmetry breaking. For our investigations, we use the nonperturbative renormalization group formalism, using a local potential approximation to construct an approximate solution of the flow. Moreover, we focus on the nearly continuous signal build as a perturbation of the Marchenko-Pastur law with many discrete spikes.
\medskip

\end{abstract}

\maketitle

\section{Introduction}
The renormalization group (RG) is one of the most important discoveries of the XX$^{th}$ in physics. It is more a general idea rather than a specific law of nature; aiming to extract relevant features of statistical or quantum states in a modern conception due to Kadanoff and Wilson \cite{KADANOFF:1967zz}-\cite{Wilson:1974mb}. Introduced in the area of statistical physics, it is, in particular, the most powerful concept to explain the universality of large distance physics for systems involving a very large number of interacting degrees of freedom, without requiring a complete description of these fundamental degrees of freedom. RG explain universality and efficiency of effective descriptions of physical laws through a progressive dilution of information with \textit{coarse-graining}, which are absorbed into the \textit{running} parameters defining effective theory \cite{Beny:2015vka}-\cite{Beny:2012qh}. The most universal formalization of the RG is based on the existence of an intrinsic hierarchy of degrees of freedom; in such a way that we can progressively ignore some of them, “integrated" in a less fundamental effective description for the remaining ones. For this reason, RG is particularly relevant in many-body physics, for all problems involving a very large number of interacting degrees of freedom. In physics, this hierarchy is intrinsically related to the notion of scale; and RG aims to construct large scale effective theory integrating out microscopic degrees of freedom, in such a way to preserve long-distance physics see \cite{Delamotte:2007pf}-\cite{Manohar:2020nzp}. More Generally, the Kadanoff and Wilson idea is the statement that the best way to study a sub-number of degrees of freedom in a large system is to integrate out the remaining degrees of freedom. Standard incarnations of the RG takes the form of a flow in the formal space of Hamiltonians (log-likelihood in probability theory), describing a sequence of distributions having the same long-distance physics.
\medskip

\begin{figure}
\begin{center}
\includegraphics[scale=0.23]{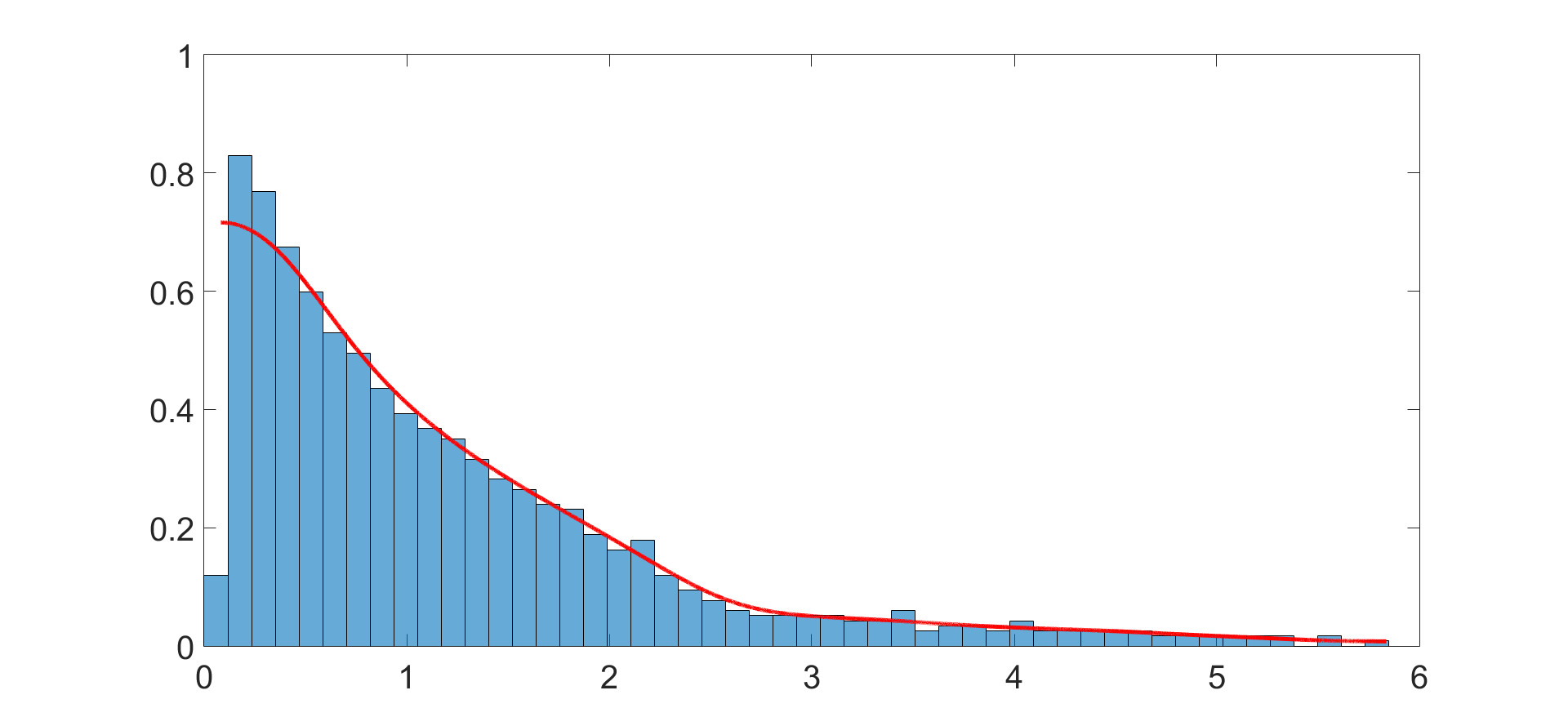}
\end{center}
\caption{A nearly continuous spectra obtained as a deviation (by adding large number of spikes) to the spectra corresponding to the covariance matrix of a random matrix with i.i.d entries. For such a spectrum standrad PCA fails to provide a clean separation between relevant and irrelevant degrees of freedom. }\label{fig1}
\end{figure}

\begin{figure}
\begin{center}
\includegraphics[scale=0.8]{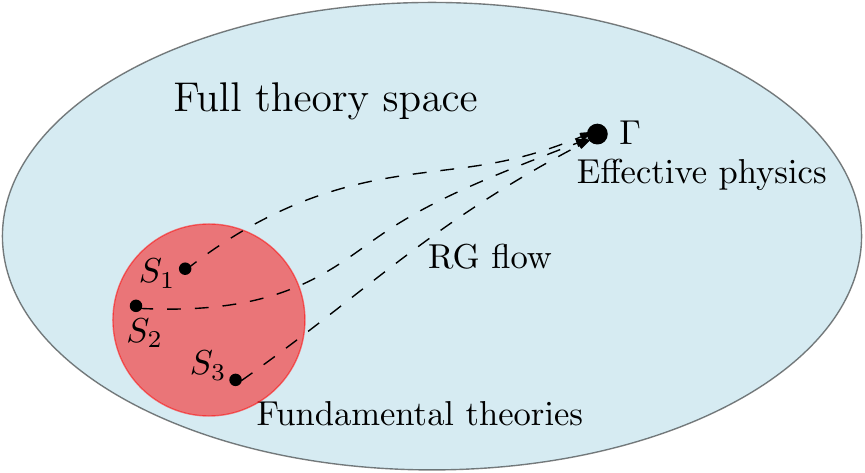}
\end{center}
\caption{RG flow through the full theory space, which is the largest set of physically relevant hamiltonians. From elementary hamiltonians $\mathcal{S}_1$, $\mathcal{S}_2$... in the red bubble, corresponding to different microscopic physics, which may converges toward the same effective description $\Gamma$. }\label{fig2}
\end{figure}

Data analysis and machine learning are aiming to extract relevant features among sets of very large dimension. This is in particular the case within the big data paradigm. Principal component analysis (PCA) \cite{Lahoche:2020oma}-\cite{pca1} look for a linear projection into a lower-dimensional space keeping only relevant features; exactly what the RG aims to do. Thus, RG is expected to be a relevant and competitive approach to standard PCA. In this paper, we focus especially on a problem where standard PCA fails to provide a clean separation between “what is relevant" and “what we can ignore". This is, in particular, the case of nearly continuous spectra, as Figure \ref{fig1} shows. For such a spectrum standard PCA does not work suitably and RG is expected to be able to provide a distinction between signal and noise. This can be achieved through a field theoretical embedding, as considered in \cite{Lahoche:2020oma}, from an analogy with what happens in standard field theory. The number of relevant terms in the Hamiltonian, spanning the distinguishable distributions at large scales, depends on the dimension of space $d$, and such, on the momenta distribution $\rho(p^2)=(p^2)^{d/2-1}$. The relevance of couplings involved in the Hamiltonian thus depends on the momentum distribution. From this basic observation, it seems to be reasonable to investigate the RG flow associated with the eigenvalue distribution of the covariance matrix through a suitable field theoretical embedding able to reproduce (at least partially) the data correlations and extract relevant features of the distributions. Note that such a strategy follows the current point of view about field theory, understood as effective descriptions at the large scale of some partially understood microscopic physics \cite{Lahoche:2020oma}-\cite{pca0}. In this way, a signal could be differentiated from noise by simple comparisons of the classes of equivalences generated by the relevant couplings. Note moreover that such a strategy does not allows \textit{à priori} to infer effective properties of data. In this paper, we only claim to build an effective theory, in the same class of long-range equivalence as the "true" theory, see Figure \ref{fig2}. This point of view was the one developed in \cite{Lahoche:2020oma}-\cite{pca0}. In these papers, the authors were able to characterize the presence of a signal, and estimate the breaking point between signal and noise, by the fact that the first non-Gaussian perturbation, which is relevant for a purely Marchenko-Pastur (MP) distribution and becomes irrelevant for a sufficiently strong signal. In this paper, we focus precisely on the asymptotic aspects (IR) attached to the signal and we show that a phase transition, corresponding to a breaking of reflection symmetry can be associated with it. We moreover justify the existence of an intrinsic detection threshold and show how this threshold could be considered for the construction of a functional detection algorithm. Finally, we mention some open questions.

\section{Framework}

We consider a set of data described by a big $N\times P$ matrix $X_{ia}$ for $i=1,2,\cdots, N$ and $a=1,2,\cdots P$. We assumes $N,P\gg 1$ but $P/N <1$. The covariance matrix $C$ is the $N\times N$ entries $C_{ij}=\sum_{a=1}^P X_{ia} X_{ja}$. Moreover, when the entries of X are purely i.i.d, the eigenvalues of the matrix $C_{ij}/N$ converge in the weak topology in distribution toward the MP law \cite{Lu:2014jua}. Figure \ref{fig1} provides a typical spectrum for $P=1500$ and $N=2000$. We denote by $\mu_{\text{exp}}(\lambda)$ the eigenvalue distribution.
\medskip

In \cite{Lahoche:2020oma}-\cite{Lahoche:2020oxg} we introduced a field theoretical embedding aiming to reproduce data correlations. The framework describes a nearly continuous random field $\varphi(p)\in \mathbb{R}$, the variables $p$ being defined such that $p^2$ is an eigenvalue of the covariance matrix translated from its largest eigenvalue $\lambda_0$. The field is provided with a probability density $p[\varphi]:=e^{-\mathcal{S}[\varphi]}$. The functional hamiltonian $\mathcal{S}$ being defined as:
\begin{equation}
\mathcal{S}[\varphi]:=\frac{1}{2}\sum_p \, \varphi(p) (p^2+m^2) \varphi(-p)+ gU[\varphi]\,.
\end{equation}
For $g=0$, the model is purely Gaussian, and the $2$-point correlations functions $\langle \varphi(p) \varphi(p^\prime) \rangle =(p^2+m^2)^{-1} \delta_{p,-p^\prime}$, where $\delta$ is the Kronecker delta and the notation $\langle X[\varphi] \rangle$ denotes the mean value of the quantity $X$ with respect to the probability measure $e^{-\mathcal{S}[\varphi]} \prod_p d\varphi(p)$. In that case, we reproduce exactly the experimental $2$-point correlations given by the eigenvalues of the covariance matrix if, firstly $m^2=1/\lambda_0$, and secondly the momenta $p$ are such that $p^2$ is distributed following the eigenvalue distribution of the covariance matrix. We denote as $\rho(p^2)$ this eigenvalue distribution inferred from the knowledge of $\mu_{\text{exp}}(\lambda)$, the integration measure for the variable $p$ reading as -- $\rho(p^2)p dp$ -- \cite{Lahoche:2020oma}-\cite{Lahoche:2020oxg}.

\medskip
The existence of $n$-points correlations functions which cannot be decomposed as a product of $2$-point functions accordingly to the Wick theorem require to remove the condition $g=0$. The functional $U[\varphi]$ is assumed to be a \textit{conservative} and $\mathbb{Z}_2$-invariant polynomial in $\varphi$ of the form:
\begin{equation}
U[\varphi]= \sum_{L=1}^M\, u_{2L} \, \delta_{0,\sum_{\alpha=1}^{2L}p_{\alpha}} \prod_{\alpha=1}^{2L} \varphi(p_\alpha)\,.\label{int}
\end{equation}
It is conservative in the usual sense in field theory, meaning that momenta are conserved at each vertex. The choice of these interactions and the reflection symmetry $\varphi\to -\varphi$ follow from simplicity. Indeed, we are aiming to construct only an approximation, and extract some relevant features concerning the momenta distributions able to discriminate between data and noise. See \cite{Bachtis:2020fly} for more details.

\medskip
The RG flow can be constructed from the standard Wilson-Kadanoff procedure, partially integrating over modes having high momenta (ultraviolet (UV) modes). In such a field framework, it is suitable to use the functional renormalization group (FRG) to construct approximate solutions of the RG flow, beyond perturbation theory \cite{Delamotte:2007pf}-\cite{Manohar:2020nzp}. The FRG is based on the effective hamiltonian for integrated modes below some scale $k$ rather than on hamiltonians for the remaining, not integrated modes above the scale $k$ (infrared (IR) modes). The effective hamiltonian for integrated degrees of freedom is denoted as $\Gamma_k[M]$ and obeys to the first order differential equation:
\begin{equation}
k \frac{d}{dk}{\Gamma}_k= \frac{1}{2}\, \sum_{p} \dot{r}_k(p^2)\left( \Gamma^{(2)}_k +r_k \right)^{-1}_{p,-p}\,. \label{Wett}
\end{equation}
In this equation:
\begin{itemize}
\item $r_k(p^2)$, the regulator, plays the role of an effective mass, depending both on momenta and infrared cut-off $k$. It vanishes for high momenta with respect to $k$ ($p^2/k^2 \gg 1$), whereas low momenta modes are frozen, and decouple from long distance physics. Moreover, $r_k(p^2)$ vanishes for $k=0$, ensuring that all the modes are integrated out.
\item The effective averaged hamiltonian $\Gamma_k[M]$ is defined from a slight modified version of the Legendre transform for free energy $W_k[j]$:
\begin{equation}
\Gamma_k[M]+W_k[j]=\sum_p j(-p) M(p)+\Delta \mathcal{S}_k[M]\,,
\end{equation}
where $\Delta \mathcal{S}_k[\varphi]:= \frac{1}{2}\sum_p \varphi(p) r_k(p^2) \varphi(-p)$. The free energy $W_k[j]$ being the generating functional of cumulants, $W_k[j]:= \ln \Big\langle \exp \left(\sum_p j(-p) \varphi(p)+\Delta \mathcal{S}_k[\varphi]\right)\Big\rangle$. This definition ensures that $\Gamma_k$ reduces to the microscopic hamiltonian $\mathcal{S}$ in the deep UV ($k^2 \gg 1$), where $r_k(p^2)$ is expected to be of order $k^2$. Moreover, for $k=0$, $r_k(p^2)$ vanishes, and $\Gamma_k$ reduces formally to the full effective hamiltonian $\Gamma$, with all modes integrated out.
\item The notation $\Gamma^{(2)}$ means second derivative with respect to $M$, the classical field defined as:
\begin{equation}
\frac{\partial W_k[j]}{\partial j(-p)}=M(p)\,.
\end{equation}
\end{itemize}
The exact flow equation \eqref{Wett} works in an infinite-dimensional space of functions, and cannot be solved exactly in general. A standard method to construct approximate solutions is to truncate into a finite dimensional subspace, assumed to be relevant from physical conditions. In this paper we focus on the local potential approximation (LPA), assuming that non quadratic part of $\Gamma_k$ may be spanned by local interactions of the form \eqref{int}. For the quadratic part, we use standard derivative expansion (DE), keeping only couplings of order $p^2$,
\begin{equation}
\Gamma_{k,\text{kin}}[M]=\frac{1}{2} \sum_{p} M(-p) (p^2+u_2)M(p)+\mathcal{O}(p^2)\,. \label{kin}
\end{equation}
Assuming to work in the IR region, and following the standard LPA assumptions, we project the flow equation on a constant classical field, neglecting its momentum dependence: $M(p)=M\delta_{p0}$. It is suitable to include the term of order $(p^2)^0$ in the non-quadratic part. Denoting it as $U_k$, we assume the following expansion around non-vanishing vacuum $\kappa$ for constant classical field:
\begin{equation}
U_k[\chi]=\frac{u_4}{2!} (\chi-\kappa)^2+\frac{u_6}{3!} (\chi-\kappa)^3+\cdots\,,
\end{equation}
Where $\chi=M^2/2$. Despite the fact that formally no dependence on the regulator is expected for the infrared limit $k=0$; the truncation procedure may introduce a spurious dependence on the regulator \cite{Pawlowski:2015mlf}. To keep control on these spurious effects, we focus on the famous Litim regulator, which has been proved to be optimal \cite{Litim:2000ci}-\cite{Litim:2001dt} and is widely used in the literature \cite{Delamotte:2007pf}:
\begin{equation}
r_k(p^2)=(k^2-p^2)\theta(k^2-p^2)\,,
\end{equation}
$\theta$ being the Heaviside step function.
\medskip

\begin{figure}
\begin{center}
\includegraphics[scale=0.6]{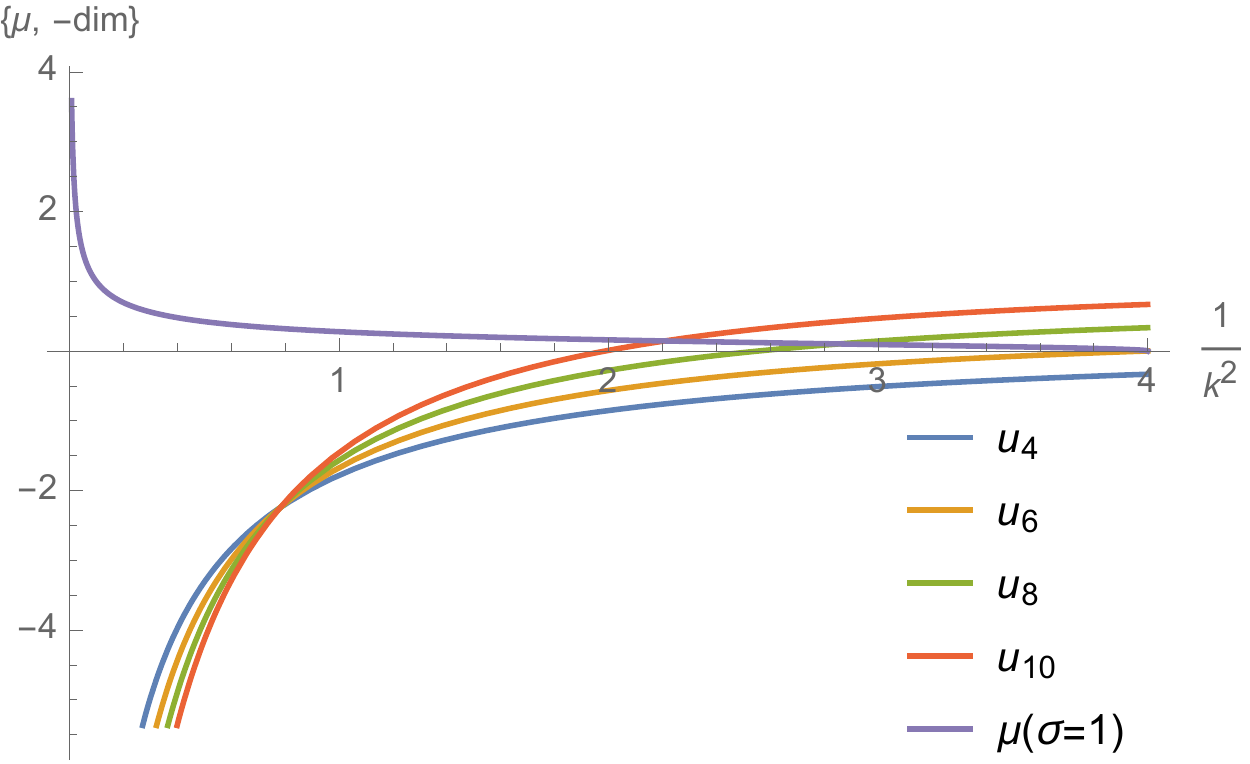}
\end{center}
\caption{Canonical dimensions for the first odd local interactions; for $\varphi^4$ (blue curve), $\varphi^6$ (orange curve), $\varphi^8$ (green curve) and $\varphi^{10}$ (red curve) asscociated to the purely MP law (purple curve) with variance equals to $1$.}\label{fig3}
\end{figure}

The flow equation for the potential $U_k$ can be deduced from the equation \eqref{Wett}, setting constant classical field:
\begin{equation}
k \frac{d}{dk}{U}_k[\chi]=\left(2 \int_0^k \rho(p^2)pdp \right)\, \frac{k^2}{k^2+\partial_\chi U_k (\chi)+ 2\chi \partial^2_{\chi}U_k(\chi)}\,.
\end{equation}
It is suitable to introduce the flow parameter $\tau:= \ln \int_0^k p\rho(p^2) dp$ rather than $k$. Moreover, from the interpretation of the parameter $u_2$ as the asymptotic effective mass, it is suitable to assume the scaling $u_2 \sim k^2$. In such a way, we are able to define a canonical dimension for all the couplings. In standard field theory, this canonical dimension allows to convert the RG equations as an autonomous system. This is not true here, because the shape of the momentum distribution is not invariant from RG transformations. However it is suitable to provide a version of dimension such that the only source of explicit scale dependence is at the level of the linear term in the flow equation. From this requirement, one expect to define dimensionless quantities denoted with a "bare" as:
\begin{equation}
\partial_{\chi}U_k (\chi)k^{-2}= \partial_{\bar\chi}\bar{U}_k (\bar{\chi})\,,\quad \chi \partial^2_{\chi}U_k(\chi) k^{-2}=\bar{\chi}\partial^2_{\bar\chi} \bar{U}_k(\bar{\chi})\,, \label{scaling1}
\end{equation}
leading to:
\begin{equation}
{U}_k^\prime[\chi]= \left(\frac{dt}{d\tau}\right)^2\, \frac{k^2\rho(k^2)}{1+\partial_{\bar\chi}\bar{U}_k (\bar{\chi})+ 2\bar{\chi}\partial^2_{\bar\chi} \bar{U}_k(\bar{\chi})}\,; \label{flowU}
\end{equation}
with the notation $X^\prime:=dX/d\tau$. We voluntary sketch the discussion on the dimension here, some details may found in \cite{Lahoche:2020oma}-\cite{Lahoche:2020oxg}. From \eqref{flowU} and \eqref{scaling1}, it is suitable to define:
\begin{equation}
{U}_k[\chi]:=\bar{U}_k[\bar{\chi}] k^2\rho(k^2) \left(\frac{dt}{d\tau}\right)^2\,, \quad \chi= \rho(k^2) \left(\frac{dt}{d\tau}\right)^2\bar{\chi}\,.
\end{equation}
The flow equation for the ‘‘dimensionless" parameter follows:
\begin{align}
\nonumber{\bar{U}}_k^\prime[\bar{\chi}]=&-\dim_\tau(U_k)\bar{U}_k[\bar{\chi}] +\dim_\tau(\chi) \bar{\chi} \frac{\partial}{\partial \bar{\chi}} \bar{U}_k[\bar{\chi}]\\
&+\, \frac{1}{1+\partial_{\bar\chi}\bar{U}_k (\bar{\chi})+ 2\bar{\chi}\partial^2_{\bar\chi} \bar{U}_k(\bar{\chi})}\,, \label{potentialflow}
\end{align}
where:
\begin{equation}
\dim_\tau(U_k)= t^\prime \frac{d}{dt} \ln \left(k^2\rho(k^2) \left( \frac{dt}{d\tau}\right)^2 \right)\,,
\end{equation}
and:
\begin{equation}
\dim_\tau(\chi)= t^\prime \frac{d}{dt} \ln \left(\rho(k^2) \left( \frac{dt}{d\tau}\right)^2 \right)\,.
\end{equation}
The flow equations for couplings $\kappa$ and $u_{2n}$ may be finally deduced from the condition:
\begin{equation}
\frac{\partial U_k}{\partial \chi}\bigg\vert_{\chi=\kappa}=0\,,\quad \frac{\partial^n U_k}{\partial \chi^n}\bigg\vert_{\chi=\kappa}=u_{2n}\,.
\end{equation}
We get for $\kappa$, $u_4$ and $u_6$:
\begin{equation}
{\bar{\kappa}}^\prime=-\dim_\tau(\chi) \bar{\kappa} +2\frac{3+2\bar{\kappa} \frac{\bar{u}_6}{\bar{u}_4}}{(1+ 2\bar{\kappa} \bar{u}_4)^2}\,,\label{kappa}
\end{equation}
\begin{align}
\nonumber{\bar{u}_4}^\prime=-\dim_\tau(u_4) \bar{u}_4&+\dim_\tau(\chi) \bar{\kappa} \bar{u}_6-\,\frac{10\bar{u}_6}{(1+2\bar{\kappa} \bar{u}_4)^2}\\
&\qquad+4\,\frac{(3\bar{u}_4+2\bar{\kappa} \bar{u}_6)^2}{(1+ 2\bar{\kappa} \bar{u}_4)^3}\,.\label{u4}
\end{align}
and
\begin{align}
\bar{u}_6^\prime=-\dim(u_6) \bar{u}_6 &-12\, \frac{(3\bar{u}_4+2\bar{\kappa}\bar{u}_6)^3}{(1+2\bar{\kappa} \bar{u}_4)^4}+40\bar{u}_6\,\frac{3\bar{u}_4+2\bar{\kappa}\bar{u}_6}{(1+ 2\bar{\kappa} \bar{u}_4)^3} \,.\label{u_6}
\end{align}
The corresponding flow equation can be deduced following the same strategy.
Figure \ref{fig3} shows the canonical dimensions for the first local interactions with the pure MP law. This picture shows the existence of two regions. For the last tier of the spectrum, only two couplings are relevant, the sixtic being asymptotically marginal in accordance to power-law counting (the MP law behaving as $\rho(p^2)\sim (p^2)^{1/2}$ for small $p$). In contrast, for the two first tiers of the distribution, the number of relevant interactions may be very large. As discussed in \cite{Lahoche:2020oma} standard methods in field theory do not work suitably in such a case. One should expect that the field theoretical approach being relevant only for the last tier of the spectrum that we call \textit{learnable region}.

\section{$\mathbb{Z}_2$-symmetry breaking and signal detection}
Besides these analytic considerations, we provide in this section the first look at a numerical investigation on a more realistic signal, as illustrated in Figure \ref{fig1}. In our experiments, we focus on the distribution of the eigenvalues for two types of covariance matrix in the regime of high dimensions (typically in our experiments we consider $P = 1500$ and $N = 2000$, which gives $K (= P/N)=0.75$). First, we consider covariance matrix associated with i.i.d random entries. The distribution of the eigenvalues of such matrix converges, for large $P$ and $N$, to the MP's law, that we interpret to be data composed entirely of noise. corrsponding to a perturbation of the case of pure noise by adding a matrix of rank $R=65$ (defining the size of the signal). In our experiments we fix the variance to one and $K = 0.75$. For such a spectrum, the learnable region is expected between $\sim 2.5$ and $\sim 3.4$, where $\varphi^4$ and $\varphi^6$ are expected to be the only relevant interactions (this is an information that one can get from the study of the canonical dimensions as illustrated in Figure 3)..

\begin{figure}[h]
\centering
\includegraphics[scale=0.0841]{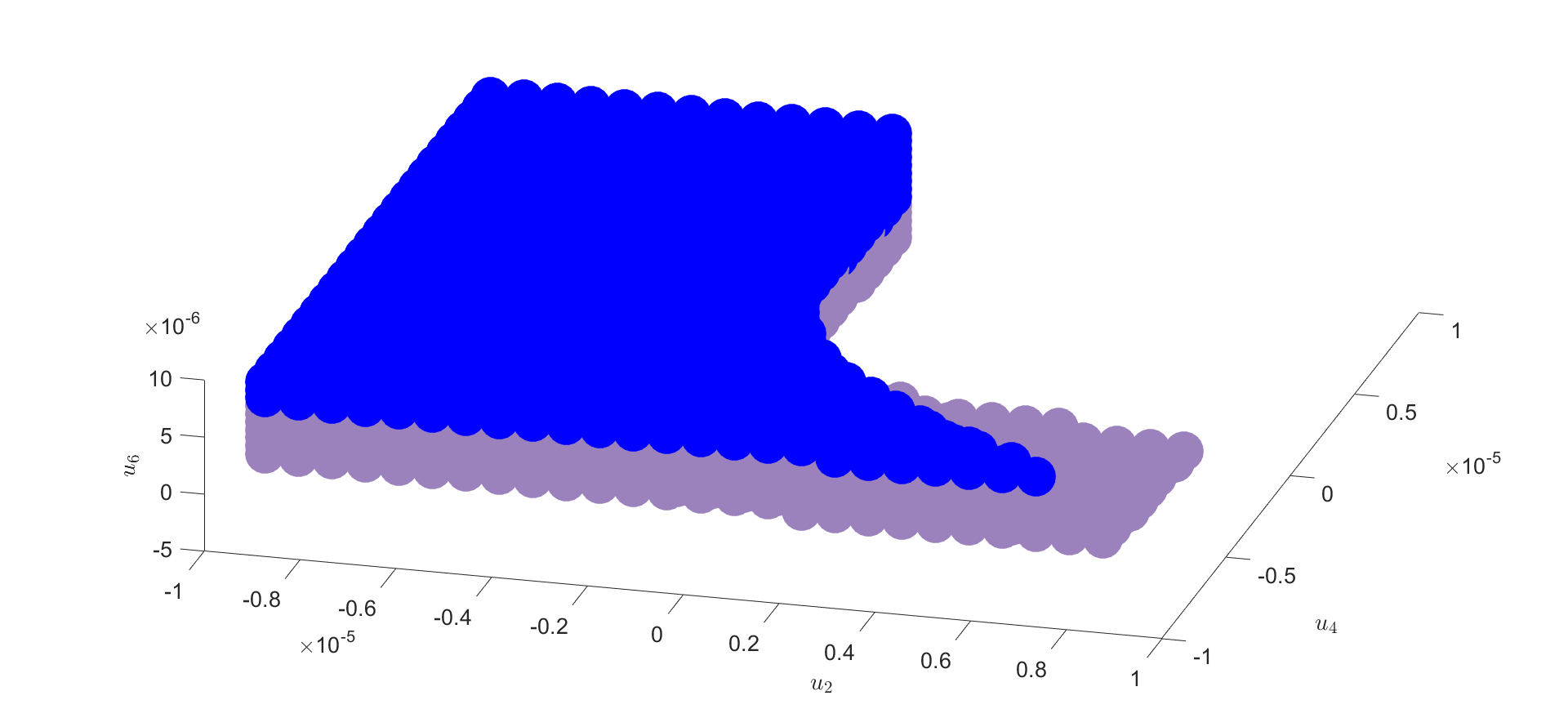}
\includegraphics[scale=0.0841]{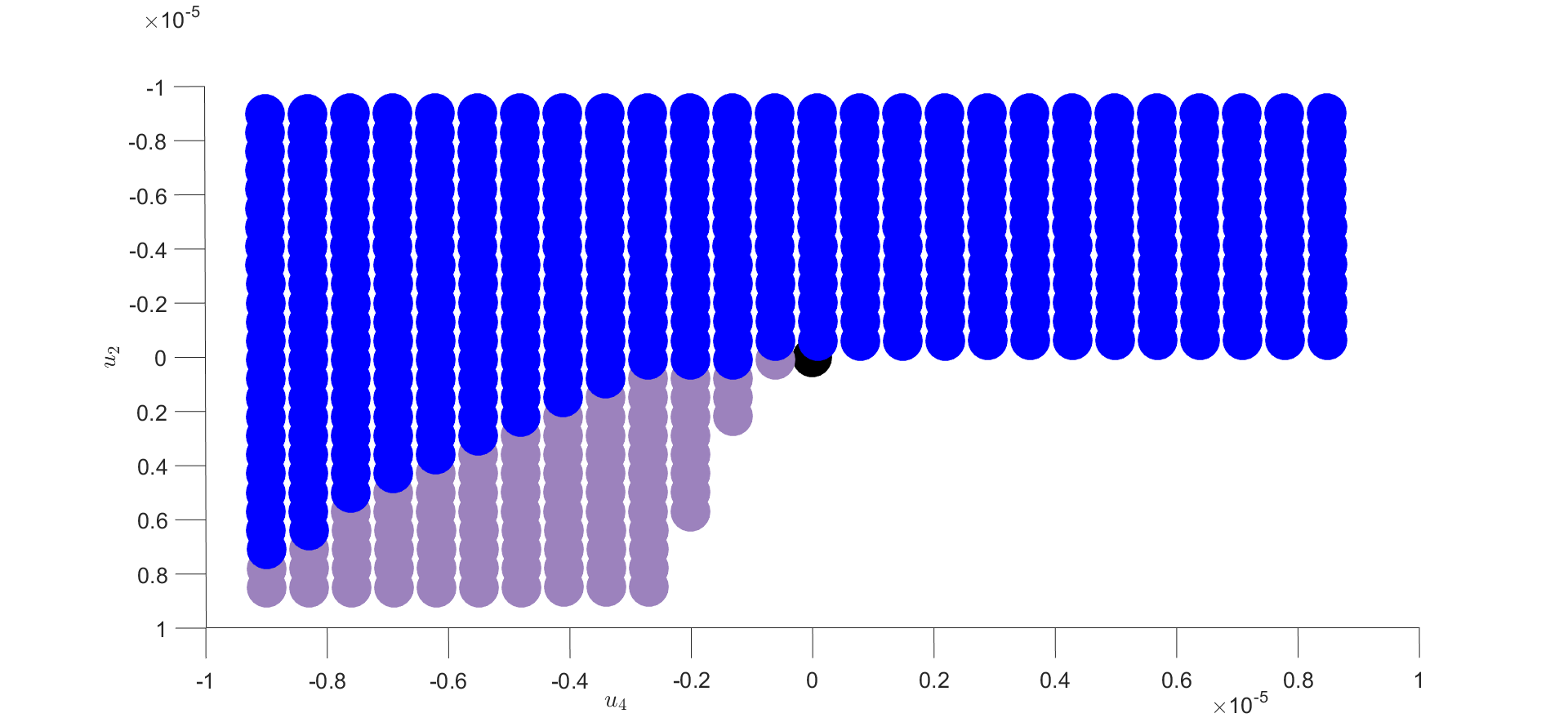}
\includegraphics[scale=0.0841]{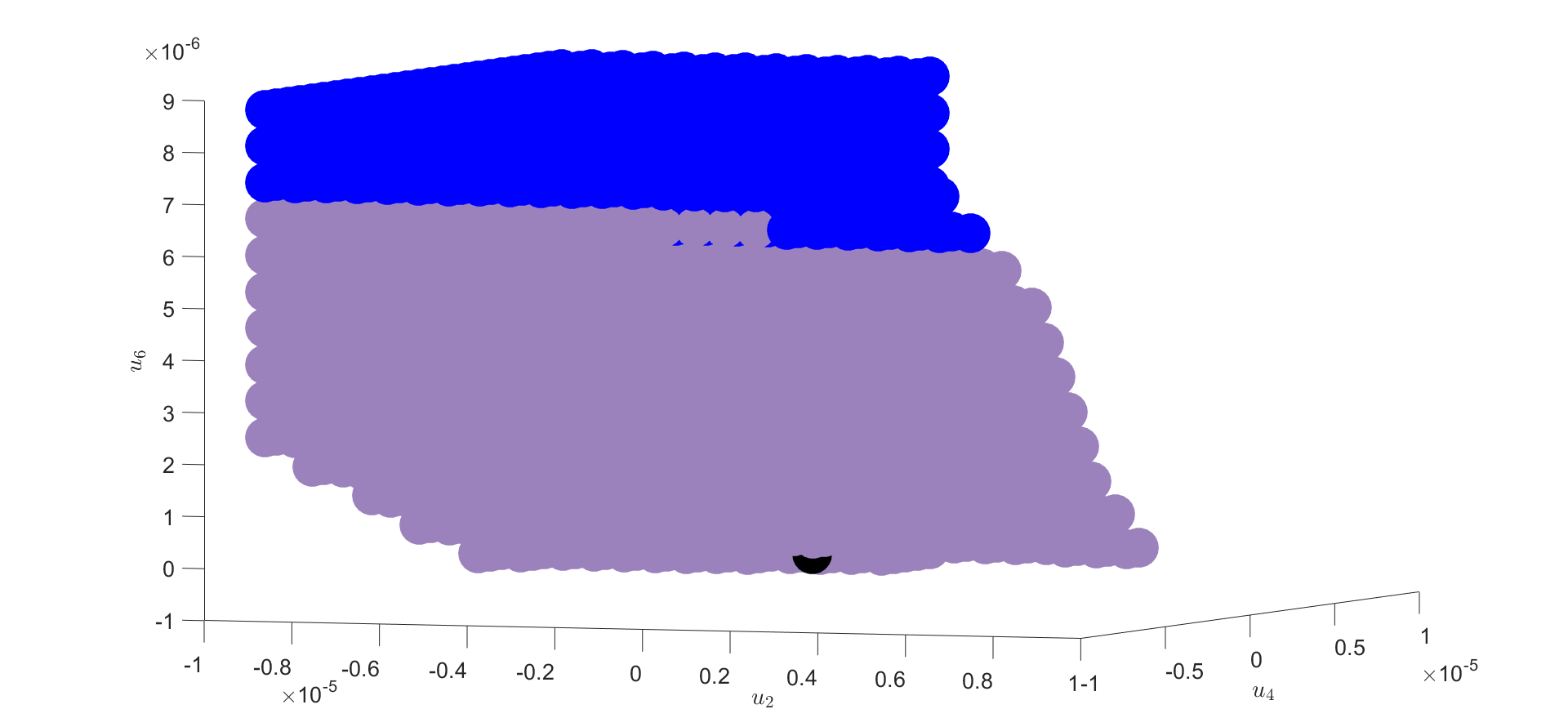}
\includegraphics[scale=0.0841]{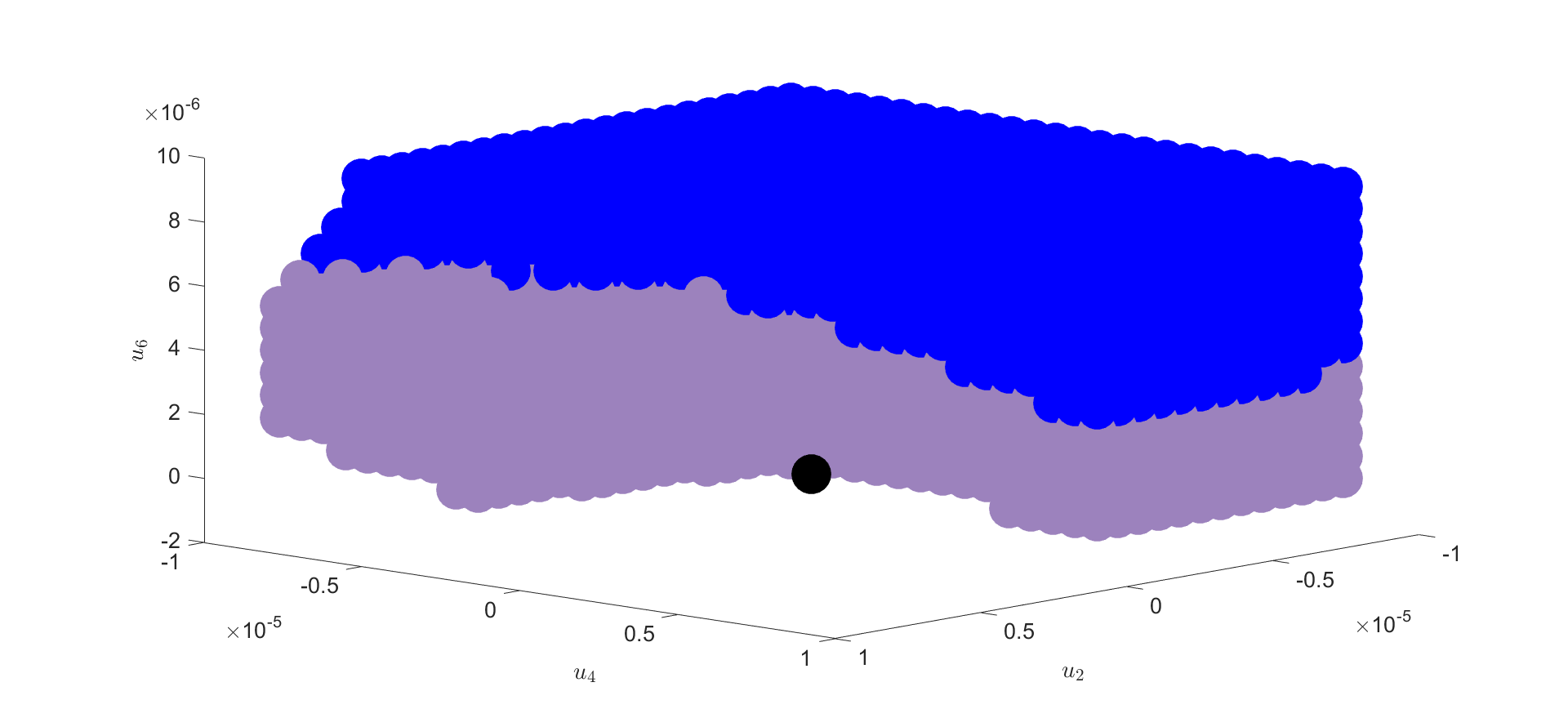}
\caption{Different view points of the compact region $\mathcal{R}_0$ (illustrated with purple dots) in the vicinity of the Gaussian fixed point (illustrated with a black dot) for the DE formalism. In this 3D region, corresponding to the case of pure noise, the RG trajectories ends in the symmetric phase, and thus are compatible with a symmetry restoration scenario for initial conditions corresponding to an explicit symmetry breaking. The blue dots correspond to RG trajectories associated to a physically relevant states in the deep infrared, i.e. the trajectories for which the values of $\bar{u}_2$ end with the same magnitude of $N=2000$.}
\label{CompactRegionR0_and_PhysicalRegion}
\end{figure}

\medskip
To start, and following \cite{Lahoche:2020oma}-\cite{Lahoche:2020oxg}, we focus on the simpler version of the derivative expansion (DE), expanding the effective potential $U_k$ as a power of $m:=M/N$:
\begin{equation}
U_k(m,\{u_{2n}\})=\frac{1}{2} u_2 m^2+\frac{u_4}{4!}\,m^4+\frac{u_6}{6!}\,m^6\,.
\end{equation}
The derivation of the corresponding flow equations follows the same strategy as for \eqref{kappa}, \eqref{u4} and \eqref{u_6}, see \cite{Lahoche:2020oma}-\cite{Lahoche:2020oxg}. In Figure \ref{CompactRegionR0_and_PhysicalRegion}, we illustrated different viewpoints of the 3D compact region $\mathcal{R}_0$ in the vicinity of the Gaussian fixed point where the RG trajectories, obtained by the DE, ends in the symmetric phase, and thus are compatible with a symmetry restoration scenario for initial conditions corresponding to an explicit symmetry breaking. However, all these initial conditions are not expected to be physically relevant in the deep IR. Indeed, for scales $k^2\sim 1/N$, one expect to obtain a good approximation for the exact covariance matrix. From construction, this imposes $u_2$ to reach a finite value, of the order of the inverse of the larger eigenvalue of the spectrum. In turn, this imposes for the dimensionless parameter $\bar{u}_2$ to be of order $N$. The initial conditions compatibles with this requirement are pictured in blue on the Figure.

\begin{figure*}
\centering
\includegraphics[scale=0.0871]{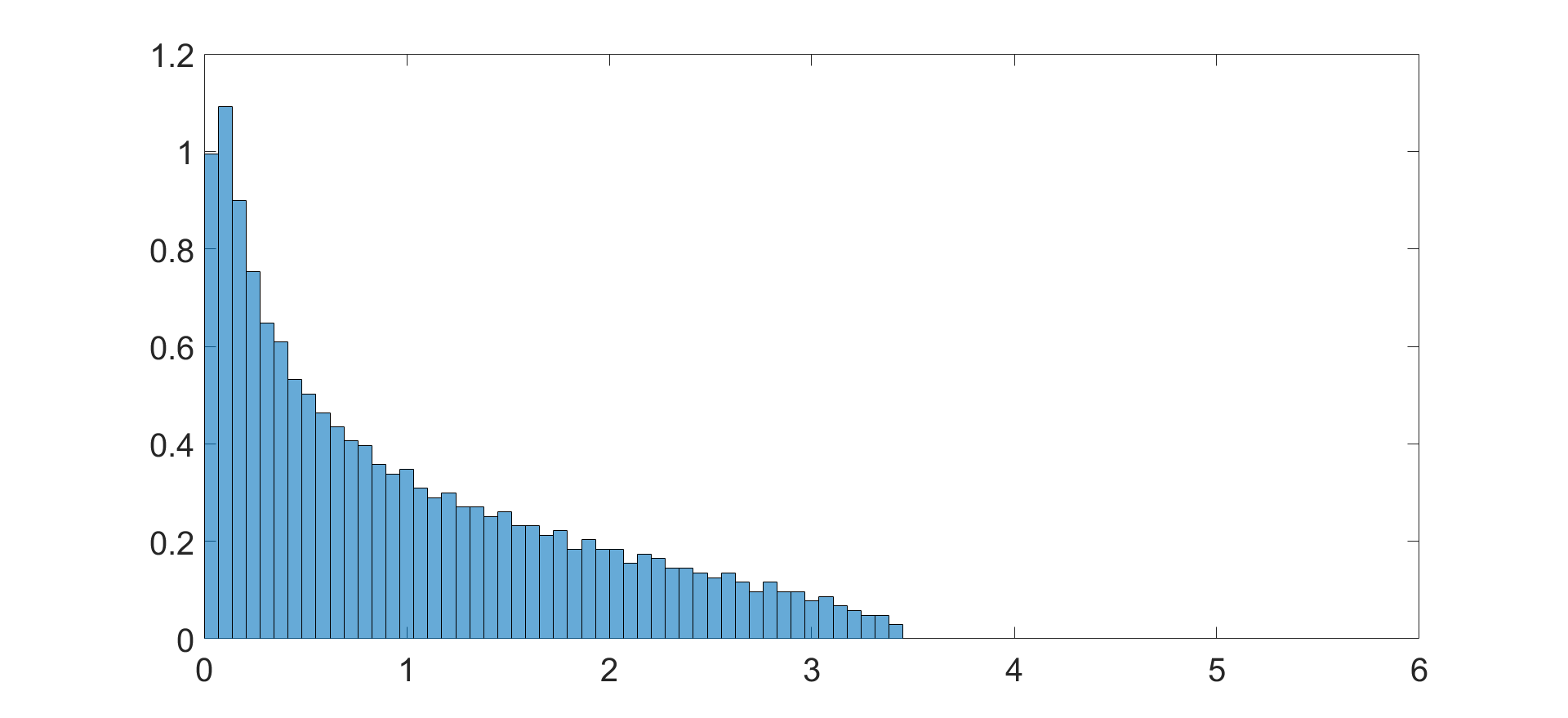}
\includegraphics[scale=0.0871]{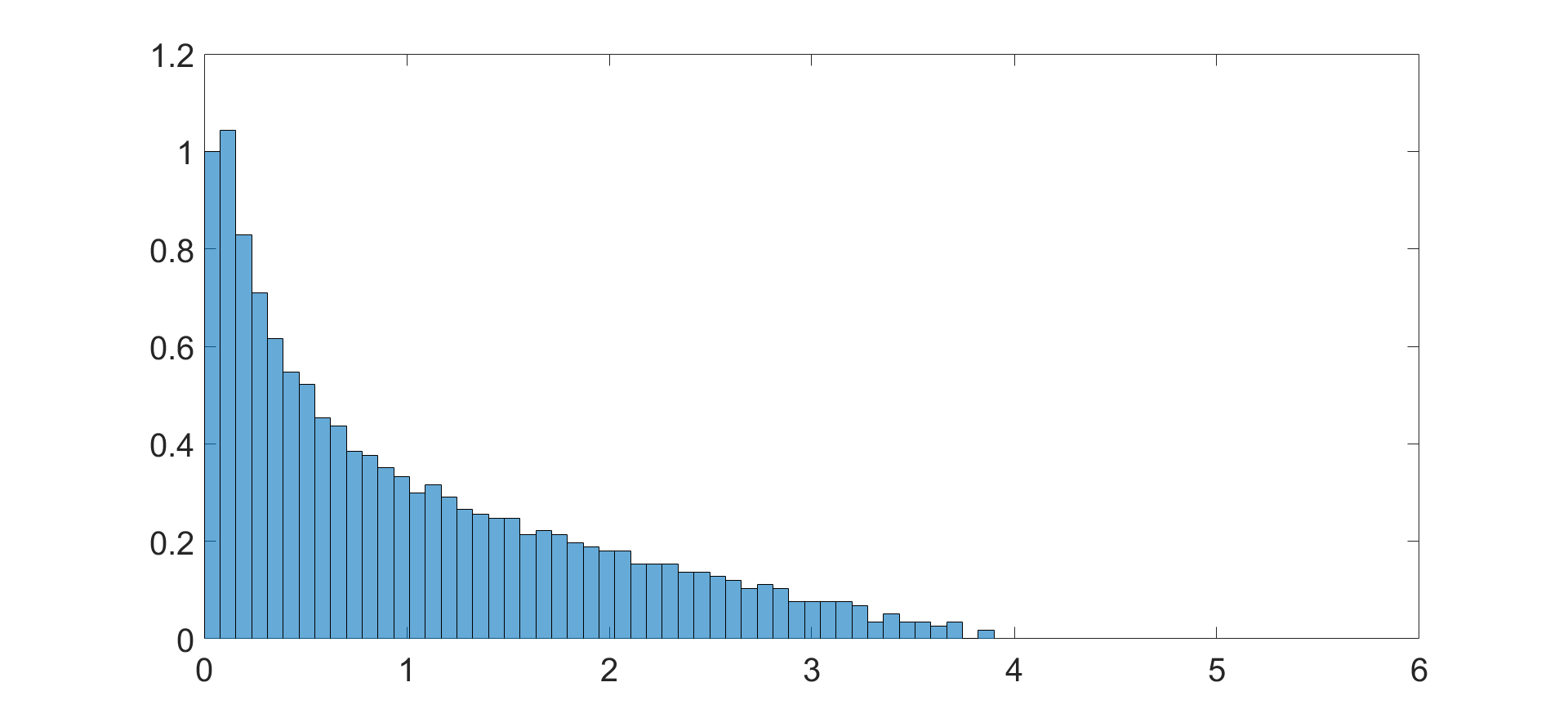}
\includegraphics[scale=0.0871]{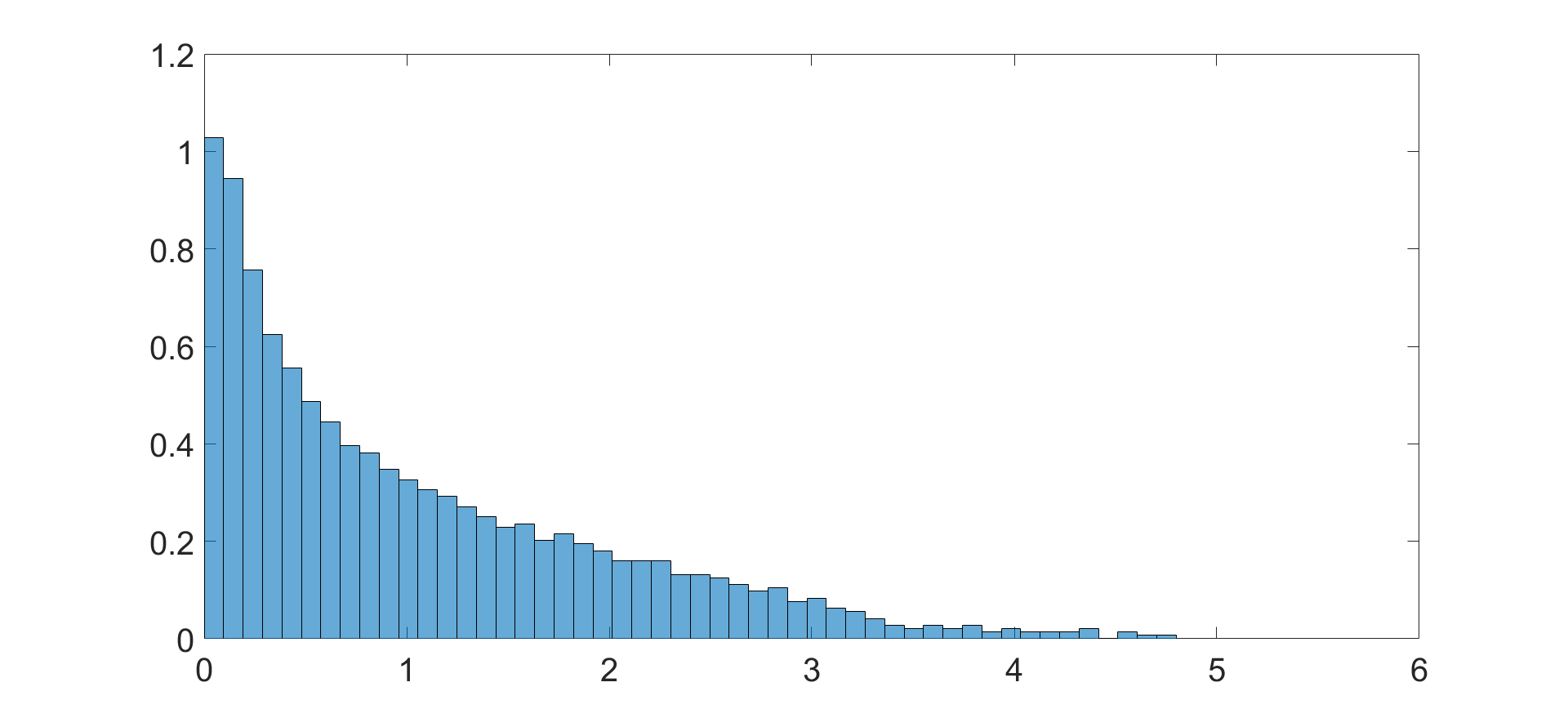}
\includegraphics[scale=0.0871]{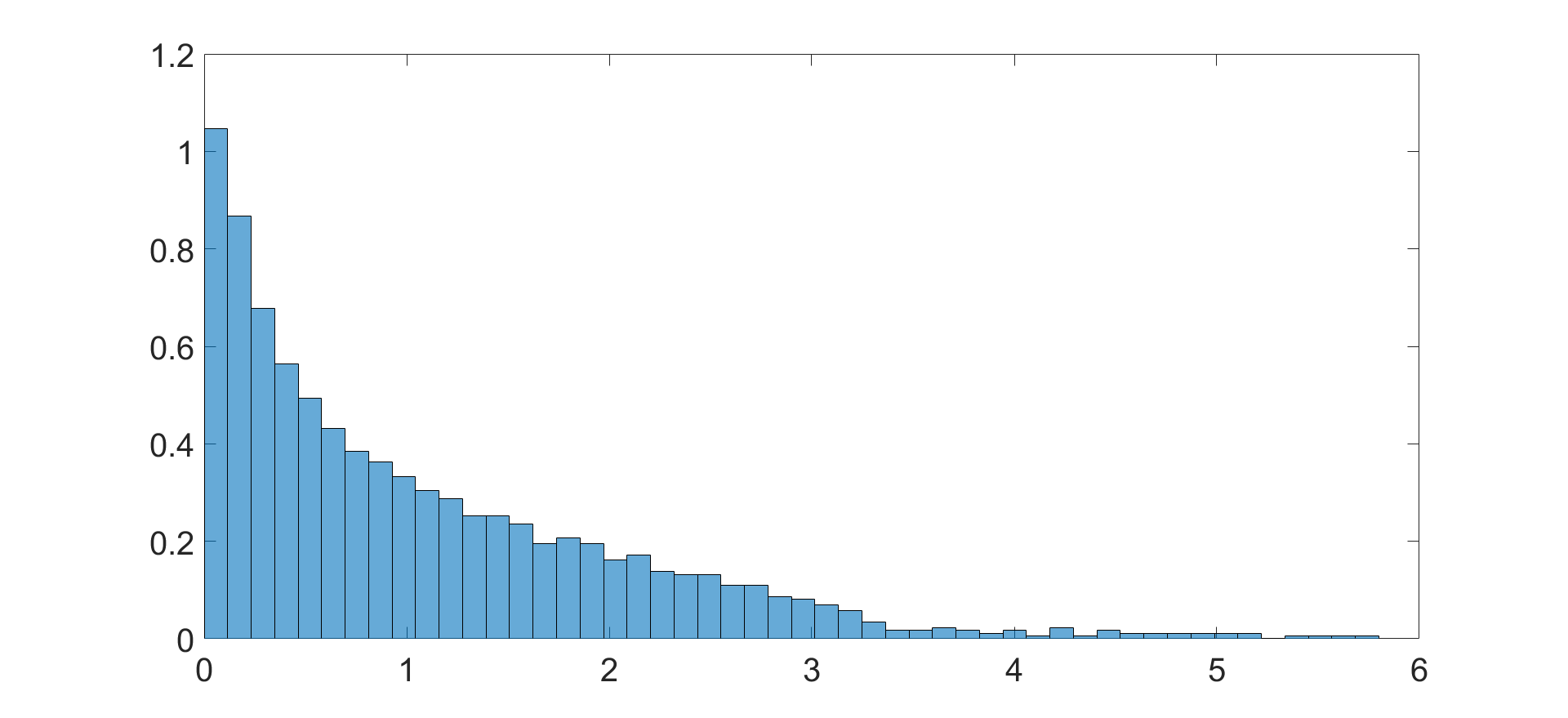}
\includegraphics[scale=0.0871]{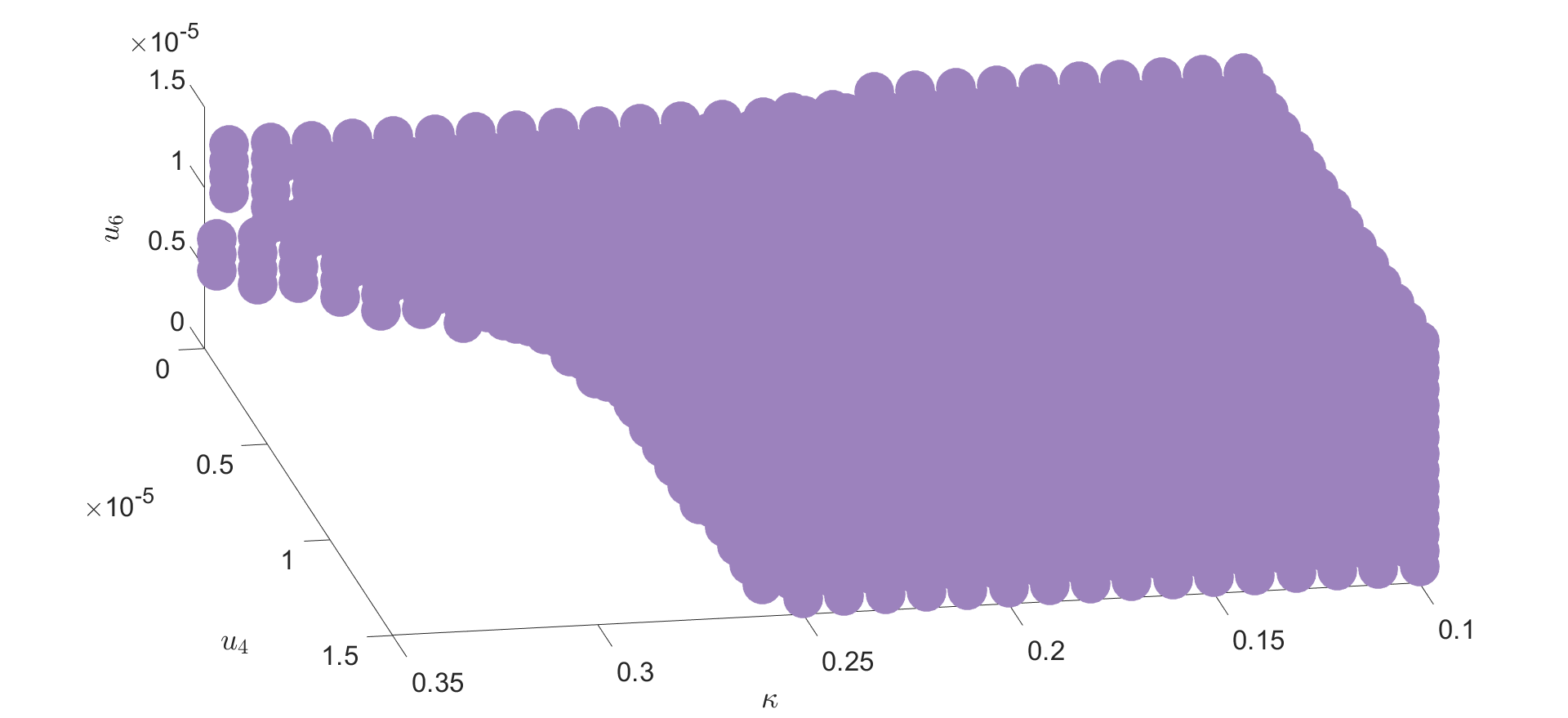}
\includegraphics[scale=0.0871]{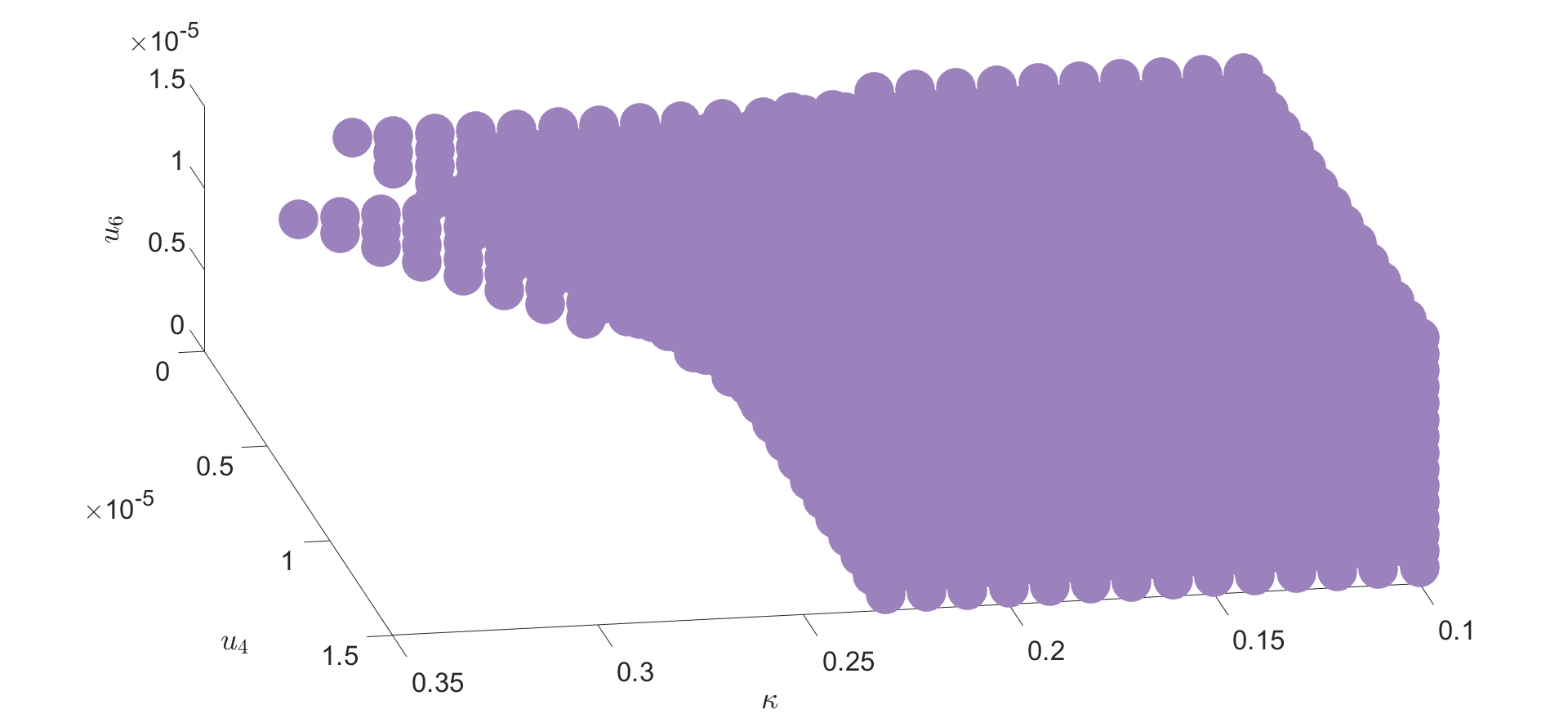}
\includegraphics[scale=0.0871]{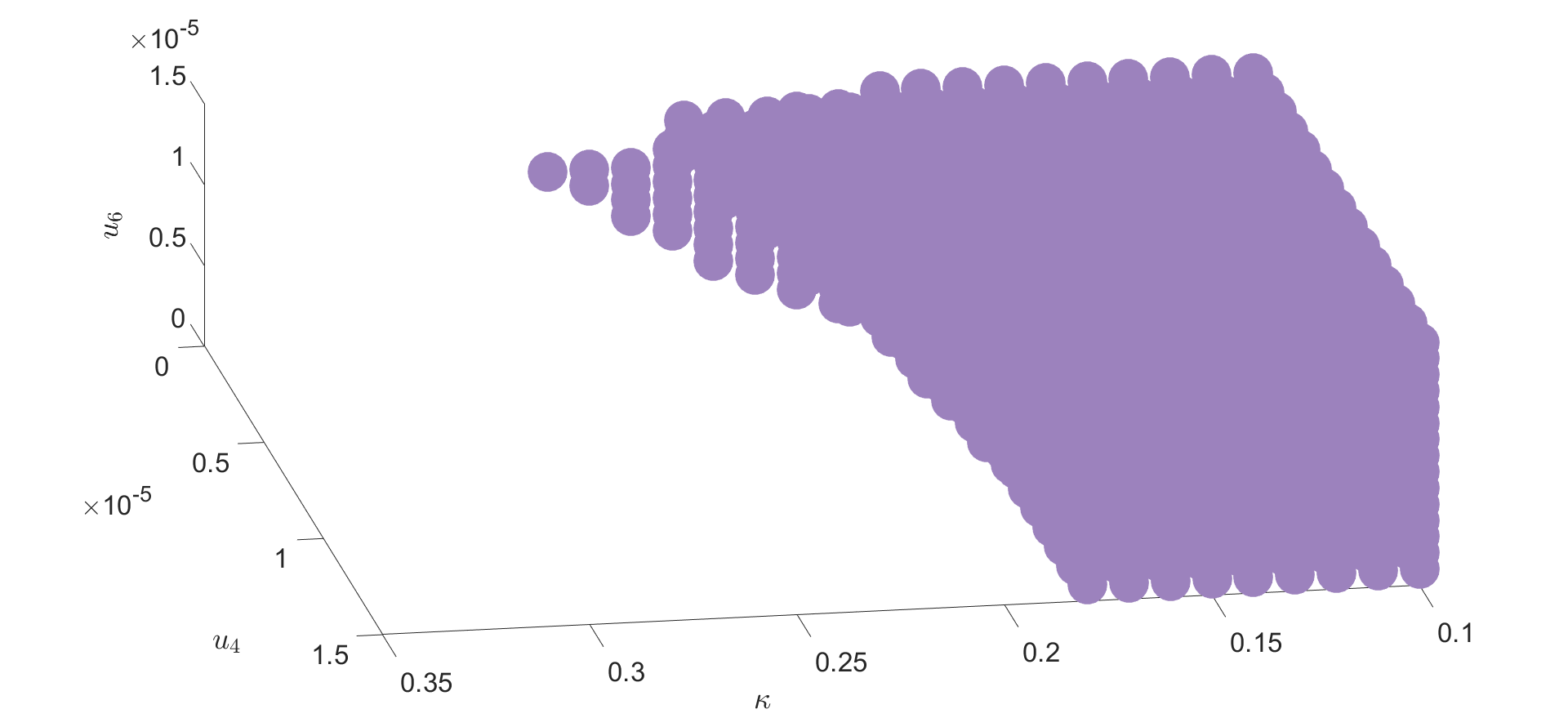}
\includegraphics[scale=0.0871]{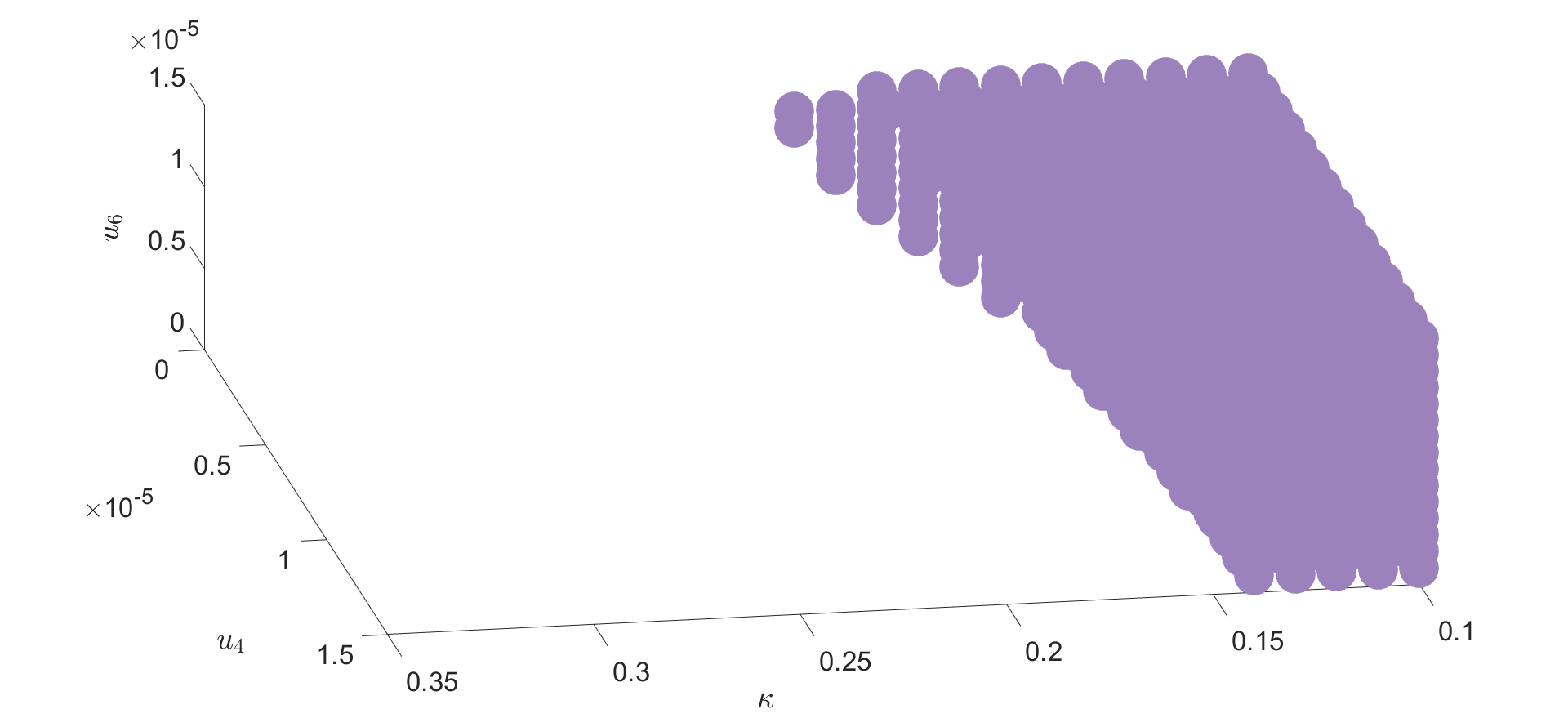}
\caption{Top: The nearly continuous spectra obtained for different intensities of the signal. From left to right we apply respectively, 0, 40, 70 and 100 percent of the intensity of the original signal. Bottom: A view point of the respective 3D compact region $\mathcal{R}_0$ (illustrated with purple dots) for the LPA formalism. In this regions, the RG trajectories ends in the symmetric phase, and thus are compatible with a symmetry restoration scenario for initial conditions corresponding to an explicit symmetry breaking.}
\label{CompactRegionR0forDifferentSignalIntensities}
\end{figure*}

\begin{figure}
\begin{center}
\includegraphics[scale=0.17]{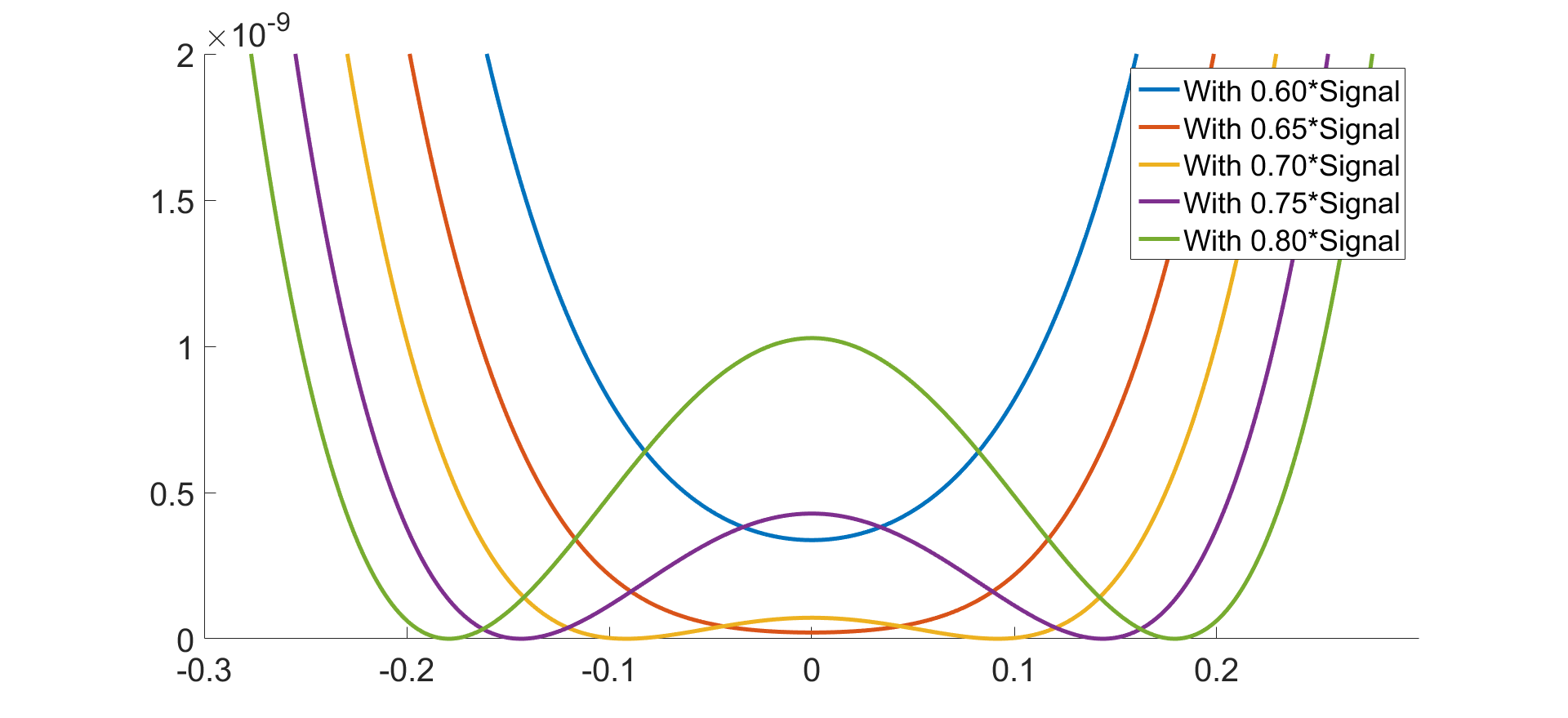}
\end{center}
\caption{Illustration of the evolution of the potential associated to a specific example of initial conditions of the coupling $\kappa$, $u_4$ and $u_6$. The different plots correspond to the potential, obtained by the LPA representation, in the deep infrared for different intensities of the signal. }\label{SymmetryBreakingRep2forDifferentSignalIntensities}
\end{figure}

In Figure \ref{CompactRegionR0forDifferentSignalIntensities} we show the same region $\mathcal{R}_0$ using LPA and equations \eqref{kappa}, \eqref{u4} and \eqref{u_6}. We show that this region is as well compact, and reduces when we increase the intensity of the signal. Finally, we illustrate on Figure \ref{SymmetryBreakingRep2forDifferentSignalIntensities} how the (deep) IR potential changes accordingly to the intensity of the signal. When the signal is low the RG trajectories end in the symmetric phase and conversely it stays in the non-symmetric phase when the signal strength is strong; providing explicit evidence of the relation between signal and symmetry breaking in the deep IR region.

\section{Conclusion}

In this paper, we investigated the RG of an effective field theory able to reproduce IR correlations at least partially in the learnable region, where both locals $\varphi^4$ and $\varphi^6$ are relevant. Focusing on local interactions, we constructed approximate solutions of the exact RG equation \eqref{Wett}, using standard DE and LPA. Some  extended discussions can be found in \cite{Lahoche:2020oma}-\cite{Lahoche:2020oxg}, especially regarding the role of the anomalous dimension, which does not change our conclusions. Among the IR properties of the effective IR theories, we focused on the vacuum expectation value. We showed the existence of a nearly compact region $\mathcal{R}_0$ in the vicinity of the Gaussian fixed point where the $\mathbb{Z}_2$-symmetry is always restored in the deep IR for purely noisy signals well described by the MP law. Furthermore, we observed that the size of this region $\mathcal{R}_0$ is reduced when we consider a deviation by a signal to the asymptotic MP spectrum. Thus, this implies that some trajectories ending in the symmetric phase for pure noise end in a broken phase, with $\langle \varphi \rangle \neq 0$ when the signal is added. Moreover, among the initial conditions allowed by the region $\mathcal{R}_0$; only a subset of them are physically relevant, i.e. such that the inverse end mass $u_2$ is of the same magnitude as the expected largest eigenvalue of the (continuous part of the) spectrum. Thus, as soon as the deformation of the region $\mathcal{R}_0$ reaches one of this physical subregion, some physically relevant trajectories are affected and leave the symmetric phase in the deep IR. This observation exhibits the existence of an intrinsic sensitivity threshold for signal detection based on the asymptotic vacuum expectation value.
\medskip

This observation allows considering a detection algorithm based on the existence of a phase transition in the deep IR region. This, however, remains an objective under investigation. In this study we focus on synthetic data for which we have a good knowledge of the noise and signal notions, essentially in order to keep control on the perturbation; but we plane to investigate this framework on real data. Other questions concern the phase transition; which seems to be able to be first or second order, depending on how the power count for the $\varphi^6$ coupling is affected. The nature of the transition could be linked to a finer detection criterion. Finally, other questions concern the approach. Investigations of regions of the more UV spectrum, for example, might require methods beyond standard DE. The validity of the field theory approximation could also be questioned in the UV. All these questions are the subject of ongoing investigations.

\medskip
To conclude, despite the fact our findings are based on a definition of noise based on the MP law, we planned to explore different mathematical incarnations of noisy signals, in a different context, to confirm the universal character of our conclusions.
Thus our investigations are also continuing for the Wigner distribution \cite{Wigner1958}, as well as on more exotic distributions for data science, based on tensors rather than on matrices and for which standard tools are also more subject to technical limitations \cite{Lahoche:2019ocf}-\cite{Lahoche:2020pjo}.

\end{document}